# Temperature-induced hysteretic behavior of resistivity and magnetoresistance of electrodeposited bismuth films for X-ray transition-edge sensor absorbers

Orlando Quaranta[1,2,*], Nunzia Coppola[3], Lisa Gades[1], Alice Galdi[3], Tejas Guruswamy[1], Alessandro Mauro[2,4], Luigi Maritato[3], Antonino Miceli[1], Sergio Pagano[2,4,5], and Carlo Barone[2,4,5]

[1]Argonne National Laboratory, 9700 S Cass Ave, Lemont, IL 60439, USA
[2]Dipartimento di Fisica "E.R. Caianiello", Università degli Studi di Salerno, Via Giovanni Paolo II 132, 84084 Fisciano (SA), Italy
[3]Dipartimento di Ingegneria Industriale, Università degli Studi di Salerno, Via Giovanni Paolo II 132, 84084 Fisciano (SA), Italy
[4]INFN Gruppo Collegato di Salerno, c/o Università degli Studi di Salerno, 84084 Fisciano (SA), Italy
[5]CNR-SPIN Salerno, c/o Università degli Studi di Salerno, 84084 Fisciano (SA), Italy
[*]corresponding author: oquaranta@anl.gov

## Abstract

This study investigates the temperature-induced hysteretic behavior of resistivity and magnetoresistance in electrodeposited bismuth films, with a focus on their application as absorbers in transition-edge sensors (TESs) for X-ray detection. Through a series of resistance versus temperature measurements from room temperature to a few Kelvins, we explore the change in the conductive behavior of bismuth electrodeposited on various substrates at the various temperatures. Our findings show for the first time both hysteretic and irreversible changes in resistivity as a function of temperature. Further, magnetoresistance measurements reveal notable variations in resistance behavior under different magnetic fields, highlighting the impact of magnetic fields on these films' electronic transport properties, with an indication of potential weak anti-localization effects at the lowest temperatures. This study not only provides a deeper understanding of bismuth's conductivity characteristics at low temperatures but also sheds light on the practical implications for developing more effective TESs for synchrotron X-ray facilities.

## Introduction

In recent years, synchrotron X-ray facilities have greatly benefited from the use of transition-edge sensors (TESs) due to their superior energy resolution compared to silicon-drift diode sensors[1]. TESs enable new science in areas such as X-ray emission spectroscopy[2-4], Compton scattering[5-7], and X-ray Absorption Fine Structure (XAFS) spectroscopy[8]. One of the primary challenges in developing X-ray TESs is to fabricate an absorber material with the necessary X-ray stopping power, while maintaining good thermalization properties and energy resolution, by controlling the total heat capacity[9]. In this regard, a promising combination of materials is gold (Au) and bismuth (Bi) due to their complementary properties. Both materials have a high atomic number (Z), which guarantee effective X-ray absorption. While Au is characterized by a well-known high thermal conductivity and relatively high specific heat[10], even at cryogenic temperatures, Bi is characterized by a very low specific heat at these temperatures, due to its semimetal nature, which strongly limits the number of available carriers in the conduction band[11]. This makes Bi very useful as an X-ray absorbing material for TESs when in combination with Au. The Au layer is necessary to ensure good thermal transport to the superconducting thermometers, while the Bi allows the formation of thick absorbers while still limiting the total heat capacity. The same semimetal nature of the Bi could represent a limiting factor for its thermal conductivity at cryogenic temperatures. Although it is possible to estimate the thermal conductivity of the Bi at cryogenic temperatures by measuring its resistivity at those temperatures (a standard approach for metals), this approach can prove challenging due to the way in which this material is typically deposited when used for TES absorbers. The standard approach for the Bi deposition in TESs is via chemical electroplating on top of an existing Au layer[12]. This approach guarantees that the Bi does not introduce artifacts in the measured X-ray spectra, typical of devices with Bi deposited via thermal evaporation[13]. The difference in behavior is reflective of the difference in crystallographic structure of the two types of films. Electrodeposited films are characterized by order of magntude larger crystals and consequently fewer grain boundaries[13,14]. The presence of an underlying Au layer makes the measurements of the Bi conduction properties challenging, especially at cryogenic temperatures, where the Au conductivity dominates. To overcome this problem, a set of 4-wire measurement devices has been developed to measure the resistivity of electrodeposited Bi while removing the contribution from the underlying Au layer. These structures provided a more accurate understanding of the electrothermal features of this material combination, which is crucial for optimizing TES performance[15]. In particular, they showed how the dependence of electrical conductivity on temperature of these electrodeposited Bi films is potentially indicative of diverse conduction phenomena such as weak anti-localization and large magnetoresistance. A better understanding of such multifaceted electrical conductivity is crucial to inform their usage as absorbers for X-rays and how this could contribute to the ultimate energy resolving capabilities of these sensors.

In this work, a series of electric and magnetic measurements performed on these devices at temperature ranging from room temperature to a few Kelvin will be described. The measurements presented have been performed on multiple devices fabricated on different substrates and measured in multiple systems, to demonstrate the validity of the data presented. Potential explanations for the phenomena seen will be discussed.

## Results

### Resistance vs Temperature

A series of resistance versus temperature $R(T)$ measurements have been collected on several devices fabricated on a variety of substrates and in multiple instruments, and the results are all consistent with each other. Examples of such measurements for microbridges of various widths, all of the same length (100 μm), deposited on different kind of substrates, are presented in Fig. 1:

- $w = 20$ μm on Hi-Res Si (green diamonds) – panel (a)

- $w = 20$ μm on AlOx (orange triangles) – panel (b)

- $w = 100$ μm on Hi-Res Si (violet squares) – panel (c)

The Bi average thicknesses ($\tau$) varied by 30 μm to 40 μm, although the thickness measurement is only indicative due to the high surface roughness caused by the large average grain size (μm)[12]. The sputtered Au thickness for all devices was 100 nm over an adhesion layer of Ti of ~ 5 nm. Starting from room temperature, the resistance exhibits a metallic temperature dependence (from here referred to as metallic), with a small reduction of the resistance from 300 K down to about 150 K. At this temperature the resistance starts to increase in a semiconductor-like fashion (from here referred to as semiconductive), although the dependence is not exactly exponential in temperature. This trend is then followed by a saturation of the resistance at temperatures below ~30 K until the base temperature of 3 K (from here referred to as plateau). Similar results have been observed in various works involving both single crystal and polycrystalline films as well as patterned devices[16,17].

The resistance also presents a series of irreversible vertical jumps, mostly during the cooldown phase. After reaching the base temperature, the resistance was measured while warming up the sample. The trajectory of the resistance proved to be similar in shape to the trajectory on the way down, but not identical. The three dependencies (metallic, semiconductive and plateau) are still present, but they develop at different temperatures. Moreover, the slope of the semiconductive phase is generally shallower on the way up, effectively reducing the extension of the metallic phase. The ending resistance at room temperature is slightly higher than where it started. The resistance is thus characterized by a combination of hysteretic behavior and irreversible status changes with temperature.

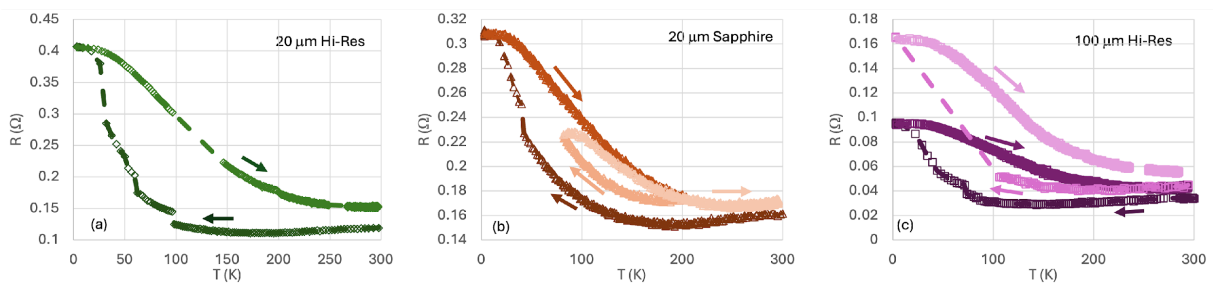

**Figure 1.** Resistance versus temperature characteristic for devices of width (**a**) 20 µm (green diamonds) and (**c**) 100 µm (violet squares) fabricated on high-resistivity silicon and of width (**b**) 20 µm (orange triangles) fabricated on sapphire. For each device, measurements collected both cooling down and warming up are presented, the arrows indicate the temperature direction (in order from darker to lighter). For the 100 µm device on Hi-Res Si and for the 20 µm device on AlOx, two consecutive sets are present.

To study this phenomenon, measurements over multiple cooldowns have been performed. In Fig. 1 two series of measurements (down and up in temperature) for all the Hi-Res Si 100 µm devices and the AlOx 20 µm device are shown. From these it is possible to see how the general dependencies in both temperature directions stay the same, but the overall resistance rises a little at every cooldown. Moreover, the specific trajectory of the resistance in temperature is dependent on the thermal history. The increase in resistance seems to slow after multiple cooldowns, and the large irreversible jumps seem to eventually disappear.

The large gaps in the data over tens of Kelvin are caused by purposely introduced interruptions in the measurements aimed to check for consistency. If the jumps in the resistance were due to artifacts in the measurement, interrupting and restarting the acquisition, which requires a long period of warming and cooling for the cryostat, we would expect discontinuities in the data sets. The absence of such discontinuities is indicative of the internal coherence of the data. Also, the fact that the jumps in resistance are always in the same direction (increase) is indicative of a real physical effect and not simply statistical noise.

To rule out the possibility that the peculiar $R(T)$ are due to the specific fabrication technique used in these devices, a set of reference devices have been fabricated. These consisted of long continuous microwires for 4-wire measurements composed of sputtered Au in some cases and of both sputtered Au and electrodeposited Bi in others. Both sets of devices were fabricated on the same wafer and were processed in the same way except for the absence of the bus current patches in Au only devices. As expected, the conductivity in the Au/Bi devices was dominated by the Au layer. By measuring both Au and Au/Bi devices at the same time it was possible to subtract one $R(T)$ from the other and obtain only the contribution of the Bi. This was characterized by the same trends shown in Fig. 1, both in terms of multiple conductivity characteristics and the conductivity hysteresis. The only difference is the absence of large jumps in resistance. This demonstrates that these multiphase and hysteretic $R(T)$ are intrinsic to the electrodeposited Bi and not due to the specific device design used here. Nonetheless, the presence of a continuous Au layer makes them not particularly useful for further studies of conduction properties.

To better understand the role of the film morphology, the 20 μm width on AlOx sample was annealed in an Argon atmosphere for about 8 hours at 150 °C. To minimize the effect of the thermal history of the sample on the resistance, i.e. to minimize the number of the large resistance jumps shown in Fig. 1, the sample was thermally cycled at least 10 times by submerging it in a bath of liquid Nitrogen and bringing it back to room temperature. Afterward, the resistivity in temperature was measured before and after the annealing, by cooling it down to about 10 K and then subsequently warming it up to 300 K. The results are presented in Fig. 2a.

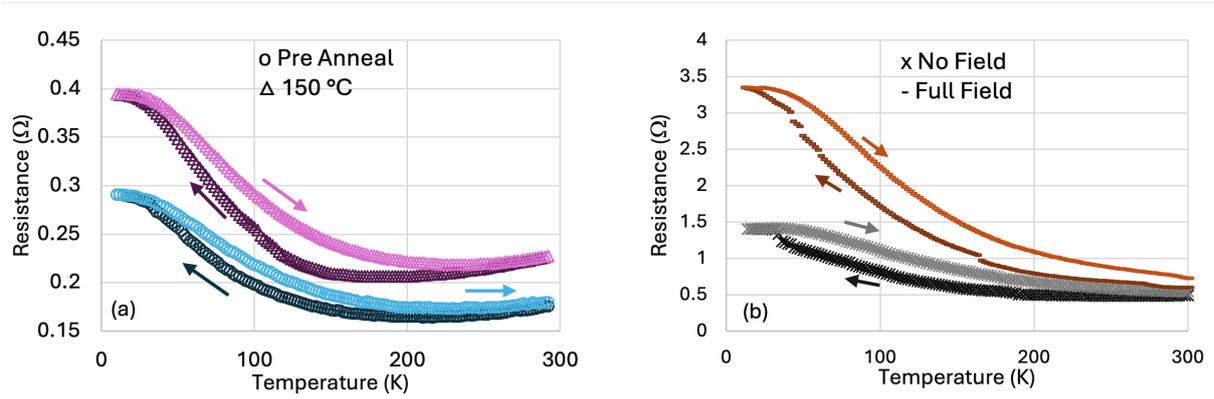

**Figure 2.** (**a**) Resistance versus temperature characteristic for an AlOx 20 μm sample. In blue circles are the data for the pristine device, while in violet asterisks are the data after annealing at 150 °C for approximately 8 hours in an Ar environment. (**b**) Resistance versus temperature characteristic for a Hi-Res Si 20 μm sample in zero field (gray x) and at 5000 G (orange -). The arrows indicate the temperature direction.

Although the general trend is the same both before (blue circles) and after annealing (violet asterisks), the details are different. Both data sets show the typical hysteretic behavior of the annealing process, but there are clear differences. Contrary to what one could expect[17], there is an overall increase in the resistance and a change in the shape. After annealing, the extension of the metallic part of the resistance is significantly increased, extending all the way down to about 150 K. This is followed by the typical semiconductive increase but at a much steeper rate. Finally, the resistance reaches the plateau below ~30 K. Another difference is in the size of the hysteresis in temperature, notably large in the sample after annealing, especially between $T$ = 75 K and $T$ = 200 K, reflective of the major difference in the extension of the metallic phase. Moreover, the difference in resistance at room temperature is smaller after annealing, indicative of a smaller permanent increase in the resistance due to thermal cycling.

## Magnetoresistance vs Temperature

To better understand the nature of the conductivity phenomena that result in the resistance versus temperature seen in the previous section, a series of magnetoresistance ($MR$) measurements have been performed. The samples were mounted in a cryostat that allowed for the measurement of resistance while a variable magnetic field ($B$) orthogonal to the sample surface was applied. The magnetoresistance, defined as the percentage variation of the resistance in field with respect to the resistance at zero field $MR(B,T) = \frac{[R(B,T) - R(0,T)]}{R(0,T)}$, was measured on multiple samples at various temperatures. The measurements covered the various conductivity regimes in temperature and were performed both during cooldown and warmup. Moreover, the magnetic field was varied between two extremes: $-0.77$ T to $0.77$ T.

The magnetic field increases the overall resistance of the device at all temperatures, but it also changes the distribution of the various phases. At full field, the metallic phase is absent, in favor of a much-extended semiconductive phase and the plateau is greatly reduced, as shown in Fig. 2b.

The magnetoresistance was measured at temperatures that cover the entire range of conductivity behavior seen in the previous section: 300 K – metallic, 180 K and 80 K – semiconductive, 13 K plateau (Fig. 3a). The resistance has a strong dependence on the magnetic field, even at these low fields. Similar trends have been seen in other works[15,16]. The two sets share the same dependence in temperature. Moreover, the sample presents a small magnetic hysteresis. When the magnetic field is swapped from one extreme (+0.77 T) to another (−0.77 T) and back, the $MR$ doesn't follow exactly the original trace, but it is slightly higher or lower depending on the sign of the field. This hysteresis grows with the lowering of the temperature, similarly to the MR behavior but on a much smaller scale, as shown in Fig. 3b (the scattering of the data at 300 K is reflective of the limited resolution on the resistance measurements and not a physical effect). This hysteresis cannot be ascribed only to the inherent hysteresis of the electromagnet used in the experiments, which is much lower that what seen here and has a dependence on the electromagnet bias current very different from what we see in our samples – maximum 0.01 T (refer to Methods sections for more details).

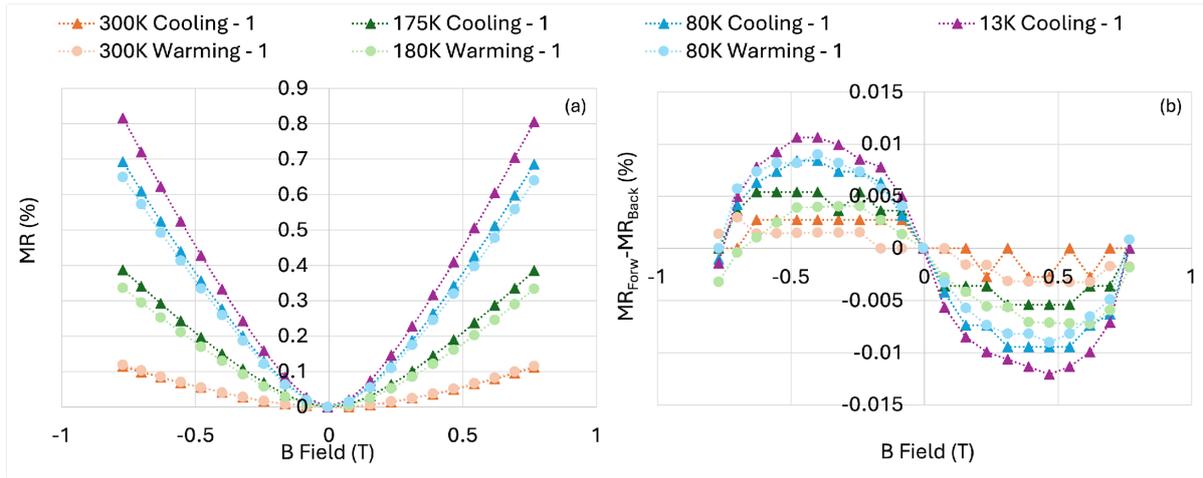

**Figure 3. (a)** $MR$ of a Hi-Res Si 20 mm sample measured at various temperatures both during cooldown (triangles) and warmup (circles). Data for each of the three phases of the resistance in temperature curve are present: 300 K – metallic, 180 K and 80 K – semiconductive, 13 K – plateau. **(b)** Hysteresis in the magnetoresistance with respect to the direction of the magnetic field sweep – from -0.77 T to 0.77 T and vice versa. The scattering of the data at 300 K is reflective of the limited resolution on the resistance measurements and not a physical effect.

Finally, the dependence of the $MR$ on the applied magnetic field is not the same at all temperatures. Classical magnetoresistance predicts a quadratic dependence at low field, followed by a saturation at high fields[18]. In Fig. 4 is represented the $MR$ in function of the magnetic field intensity squared ($B^2$) in markers and dashed lines measured at various temperatures, both while cooling down (a) and while warming up (b). In the same figure are also shown the fits to a linear dependence of $MR$ on $B^2$ using only the higher field points in solid lines. If the typical quadratic dependence was maintained, the data should lie on a straight line for all values of $B^2$. Figure 4 shows how the data follow a $B^2$ dependence only towards the higher fields. Moreover, the further down in temperature the more the dependence moves away from $B^2$, with data at the lowest temperatures having a steeper dependence on $B$. Similar behavior has been observed in other works[19,20] although at higher fields and lower temperatures.

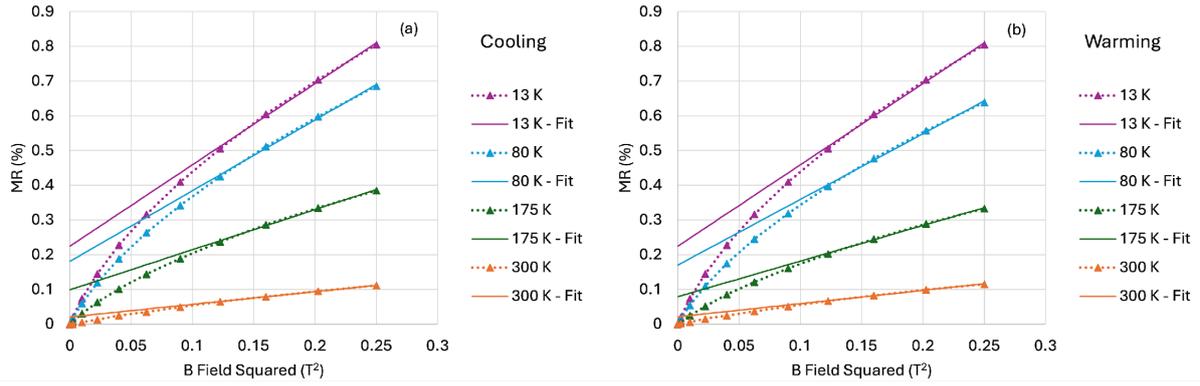

**Figure 4.** Magnetoresistance in function of field at various temperatures, measured during cooldown (**a**) and warmup (**b**). In both, data sets are represented with points + dashed lines.

## Discussion

The conduction properties of Bi films have been the subject of extensive studies throughout the years, both in continuous films[19] and in patterned structures[20]. Bi wires have been widely investigated due to the unique electronic properties of this element[21,22]. Being a semimetal with a very small indirect band-overlap (38 meV at 0 K), Bi has a temperature-dependent electron density $n_e$ ($3 \times 10^{18}$ cm$^{-3}$ at 300 K, $3 \times 10^{17}$ cm$^{-3}$ at 4 K), which is four to five orders of magnitude lower than that of ordinary metals. The Fermi surface of Bi consists of ellipsoidal pockets for electrons and holes, resulting in small effective masses $m_{eff}$ (as low as 0.001 $m_e$ – electron mass $m_e$) and large Fermi wavelengths $\lambda_F$ (~40 nm)[23]. Moreover, mean free paths $l_e$ are as long as ~100 nm at 300 K and ~400 µm at 4 K. These characteristics, together with the phonon scattering, cause the mobility to depend on temperature. A non-monotonic resistance over temperature behavior can occur, which is usually attributed to an additional limitation of the mean free path by temperature independent boundary scattering[22,23].

The resistance as a function of temperature curves shown in Fig. 1 are common in microstructures of electrodeposited Bi[17] and can be partially explained by the interplay of the temperature dependence of the electrical carrier density and of the electron mean free path. In semimetals, the Fermi temperature may be of the order of 100 K[24], which results in a limited carrier density at low temperatures with a notable temperature dependence. At the same time, the mean free path increases with decreasing temperature and can be very long in Bi at cryogenic temperatures. These two competing effects can explain the peculiar shape of the $R(T)$, that at room temperature starts with a metallic behavior, as in a regular metal, where $l_e$ increases with a decreasing temperature, while $n_e$ remains essentially constant. As the temperature decreases, approaching the Fermi temperature, $n_e$ will start to reduce. When this reduction becomes faster than the increase in $l_e$ and when $l_e$ becomes limited by the average dimension of the grain, the resistance stops behaving like a metal (where $n_e$ would remain essentially constant) and starts to behave like a semiconductor, with an increase in resistance. This increase is not precisely semiconductive in nature. In a semiconductor/insulator, the electron density of states has an exponential dependence on temperature $n_e(T) \propto e^{\frac{E_G}{2k_B T}}$, where $E_G$ is the energy gap and $k_B$ is the Boltzmann constant, which would generate a corresponding exponential dependence in $R(T)$, not present in our data. Further reducing the temperature generates a plateau in the resistance. This could (partially) be explained by the finite $n_e$ at these temperatures (typical of semi-metal, unlike semiconductors) and by the saturation of $l_e$ induced by the average crystal dimension. The presence of surface metallic states, which can dominate the conductivity at these temperatures, can also explain the saturation effect of the resistance[24]. Although these phenomena have been seen in multiple works on Bi microstructures (films and wire), the data presented in this work present some unique characteristics.

The $R(T)$ curves show the presence of irreversible jumps in the resistance, which tend to be more common during the first cooldowns and tend to disappear after several thermal cycles. These jumps in $R$ could be explained by mechanical reallocation of the crystals in the microwire, especially in the areas where the various crystals of Bi growing on top of Au surfaces merge through lateral growth in the areas where the Au is not present, typical of these 4-wire devices. These changes are due to a coefficient of thermal expansion in Bi several times higher than that of the substrates and of the Au[24]; such mismatch between the film and the substrate could lead to mechanical stress when devices are cooled. The faster contraction of the Bi crystals with respect to the substrate could increase the spacing between crystals in the microwire, with a consequent increase in the potential energy between grains, which would increase the measured resistance. Conversely, this hypothesis does not explain why the modification should be permanent. Another hypothesis is the formation of defects in the single crystals during the cooldowns. These defects can effectively reduce the average dimension of the crystals, further limiting the electron mean free path at cryogenic temperatures. This would cause an irreversible sudden increase in resistance, like the ones seen in our data. We also tried to anneal our sample at ~150 °C for 8 hours in an atmosphere of Ar. The resistance overall increased, but the metallic phase extended to lower temperatures (Fig. 2a). This partially differs from what other groups have seen[16], where they have observed a notable reduction in the resistance, together with a general change in the shape of the $R(T)$. In this work, the original electrodeposited samples show $R(T)$ very similar to the ones reported in Fig. 1, while after a suitable annealing (268 °C for 6 hours in Ar) the polycrystalline films become single crystals with the trigonal axis orientated perpendicular to the film plane. In these annealed films, the metallic phase greatly extends, essentially eliminating the semiconductive and the plateau sections. The effect is explained in terms of reduction (or effective disappearance) of grain boundary scattering. In our devices, performing the annealing at a much lower temperature (150 °C) appears to have not allowed for a complete reformation of the crystallographic structure of the film, even though these temperatures were expected to be sufficient[25]. This temperature was chosen because it would be the maximum temperature compatible with a TES fabrication process.

A possible explanation for the overall increase in the resistance could be the inelastic deformation of the film, caused by the relatively high annealing temperature without any photolithographic mold to constrain the shape (the annealing was performed on the complete device), which could have reduced the cross section in the area between the two voltage leads. Ultimately, the objective of this anneal was to test the stability of the film in real processing conditions that an electrodeposited Bi film could encounter during the fabrication of a TES. In this sense, this test proves that one must be very careful when performing these annealings because they can negatively affect the thermal conductivity of the Bi absorbers.

Another peculiarity present in our data, not reported elsewhere (to the best of our knowledge), is the hysteretic behavior of $R(T)$. The general shape is the same both cooling down and warming up, with all the various phases present: metallic, semiconductive and plateau, but with different extension. This indicates that the conduction phenomena are the same in direction but influenced by some hysteretic effect. The effect cannot be due to poor thermalization of the devices under test because the data were collected over several hours in both directions. The other hypothesis is this could be due to some kind of magnetic effect. Bi is known to be a material characterized by strong magnetic field dependent electrical conduction properties: from giant magnetoresistance[16], to Spin-Orbit Kondo effects[26], to weak anti-localization[20]. Our measurements do indeed show a reasonably strong magnetoresistance (Fig. 4), although smaller than what has been measured in single crystal Bi or in annealed electrodeposited Bi films[16,17]. In Fig. 2b we can see how the $R(T)$ is strongly influenced by the applied orthogonal magnetic field. Not only does the overall resistance increase, but also the shape of the $R(T)$ changes. In particular, the magnetic field completely removes the metallic phase at the temperatures of the experiment in favor of the semiconductive one. This is the same result shown by other groups and can be explained by the interplay between the magnetoresistance and the strong increase in conductor mean path in Bi at cryogenic temperatures[16,17]. Even at full field, the R(T) is hysteretic in temperature, similarly to what we have seen at zero field. The $MR$ measurement also shows a hysteresis in the magnetic field (Fig. 3b), which disappears at zero fields, which exclude effects of residual magnetization. This large $MR$ represents an added reason for being extremely careful with respect to residual magnetic fields when working with TESs.

Another interesting aspect of the $MR$ in these films is that it not only grows with reducing temperatures, as expected due to the increase in the conductor mean free path, but it also changes shape. The lower the temperature the further the $MR$ dependence on $B$ is far from the standard $B^2$ dependence, as shown by the quality of the quadratic fits shown in Fig. 4. At base temperature the $MR(B)$ shows a pronounced dip at low $B$, typical of weak anti-localization. In disordered two-dimensional systems, electrons - due to their wave-like nature - can interfere with themselves. This interference can double the probability of an electron returning to its starting point, which typically suppresses the conductance, a phenomenon called weak localization[27–29]. However, when strong spin-orbit coupling is present, the electron spins rotate in opposite directions, leading to destructive interference, which consequently increases the conductance, a phenomenon called weak anti-localization[30,31]. Weak anti-localization is usually challenging to identify via magnetoresistance measurements because the effect can be overshadowed by the large classical magnetoresistance. Typically to achieve a clear indication of weak anti-localization very thin films are needed, which strongly reduce the electron mean free path and consequently the classical magnetoresistance effect. Moreover, the effect is more evident at extreme cryogenic temperatures. Weak anti-localization effects have been shown in Bi thin films[20]. Our experimental setup doesn't allow us to reach temperatures below ~13 K and our samples are intrinsically thicker and composed of larger grains than those studied in Ref.[20], but nonetheless in our $MR(B)$ measures there are hints of this effect. The same weak anti-localization effect could explain the presence of the plateau at low temperatures.

## Conclusion

In this work, the electrical and magnetic conduction properties of electrodeposited Bi microbridges have been studied under various bias, temperature and magnetic field conditions. The data provide valuable information towards the understanding of the conduction properties at sub-K temperatures in electrodeposited Bi, crucial to the correct design of effective TESs for X-rays. The typical large grain structure present in electrodeposited Bi films, although it represents a strong advantage when used as absorbing material in TESs for X-rays, it is more susceptible to magnetic fields, especially at cryogenics temperatures were a strong magnetoresistance dependance to low fields is present. The increased resistance could affect the thermalization properties of the absorber, hence particular care needs to be used in shielding these devices from stray magnetic fields.

Further studies focused on quantum effects at very low temperatures like weak anti-localization and Kondo[26] in this material, possibly employing alternative experimental techniques, like noise spectroscopy[32,33], would be useful. This could shine a brighter light on Bi conduction properties at these temperatures, which is of great interest to the single photon detectors community.

## Methods

### Resistance vs Temperature

The devices used in these experiments have been discussed extensively elsewhere[15], but their nature can be summarized as following. The devices are typical 4-wire resistivity measurement structures, built around a microwire structure, represented in Fig. 5.

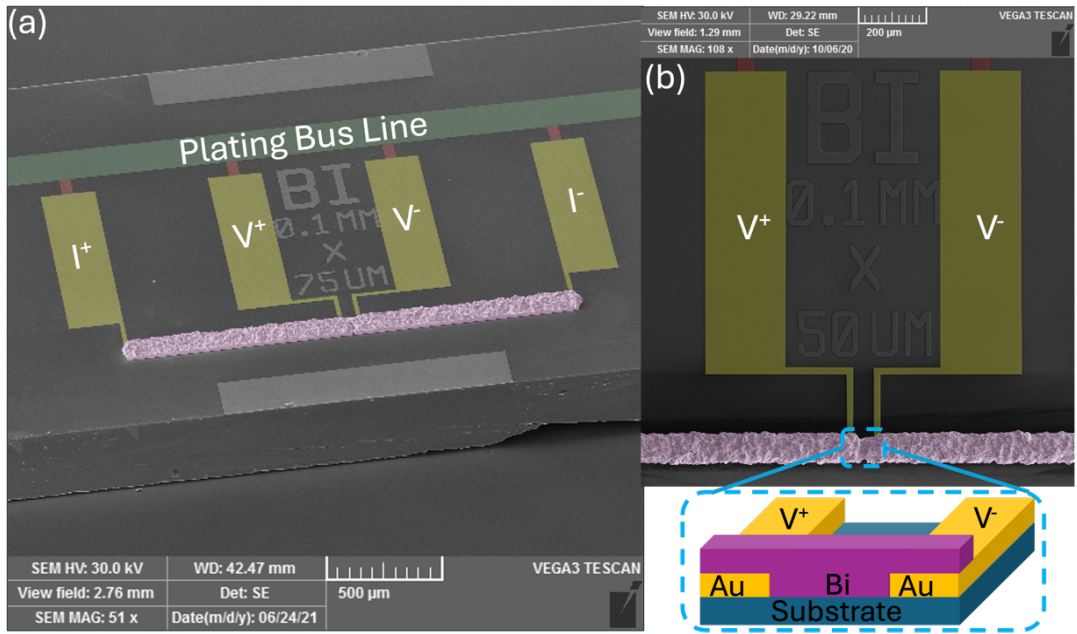

**Figure 5.** (**a**) Scanning Electron Microscope (SEM) picture of a representative device. In yellow are visible the pads for both the bias and the voltage measurements. The pads are made of sputtered Au. In violet is showed the Bi microbridge under study. Under it is also present the layer of Au that acts as seed layer for the Bi plating (not visible). In green is represented the bus line used for the plating of the Bi and in red are the connection patches to the bus line, which are removed at the end of the fabrication process. (**b**) A close-up view and a schematic representation of the Bi microbridge. The conductivity of Bi is measured between the two voltage contacts where only Bi is present (no Au seed layer).

The Bi film is grown via electroplating on top of a sputtered Au seed layer[12]. The 4-wire structure is achieved via a series of masks that define the shape of the various layers: a lift-off mask for the Au seed layer, together with the plating bus line and the connection patches, and a plating mask for the Bi microbridge. The Bi mask is shaped to allow the growth of the Bi from the Au lines, which also act as seed layers. The growth of the Bi in the plating solution is isotropic in every direction, therefore it will not only grow vertically but also horizontally, within the boundaries of the mold defined by the lithography. The Bi layers growing from the Au leads will meet in the area between the leads, over the bare substrate. This creates a microbridge between the Au lines composed by only Bi, without any Au underneath. This guarantees that measured properties are relative to the Bi and not influenced by the Au seed layer. The devices were fabricated on a variety of substrates: high-resistivity silicon (Hi-Res Si), silicon oxide coated silicon (SiOx) and sapphire (AlOx), all of which are insulators, that the electrical measurements are not influenced by them. In these structures, the two external contacts serve as the input and output lines for the bias current and the two internal are used to measure the voltage drop across the microbridge.

The microscopic structure of the Bi in the microbridge appears identical to the typical structure that is found in electrodeposited Bi layers used as X-ray absorbers in TESs. An example of both type of structures is shown in Fig. 6.

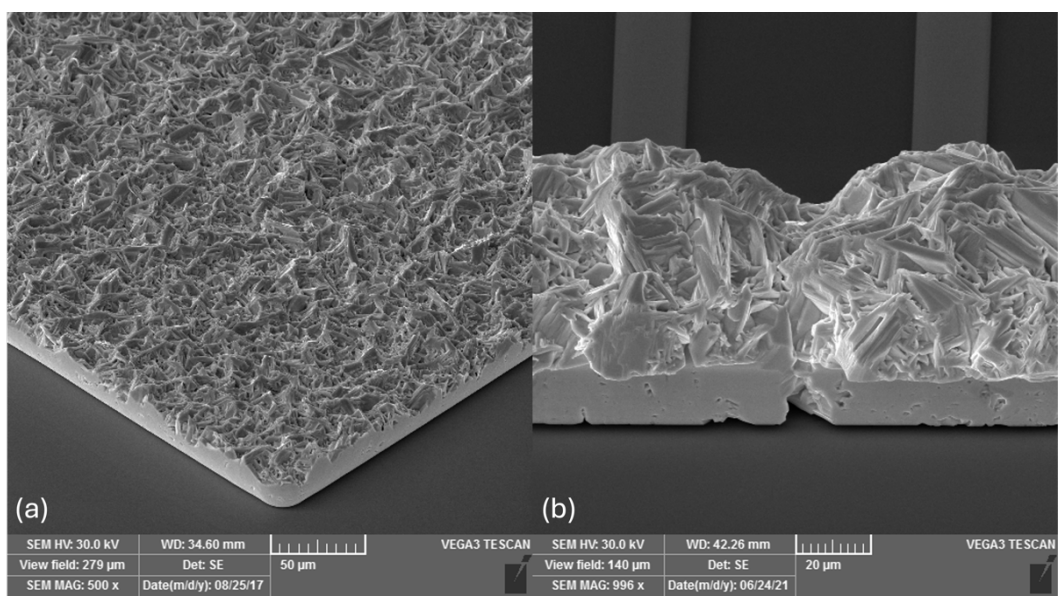

**Figure 6. (a)** Scanning Electron Microscope (SEM) picture of a representative Bi film as an X-ray absorber for TESs. **(b)** Scanning Electron Microscope (SEM) picture of a representative Bi microbridge for conductivity measurements.

## Measurements setup

The zero magnetic field DC electrical transport measurements were performed in a cryogen free Adiabatic Demagnetization Refrigerator (ADR) from FormFactor, equipped with a variable temperature stage capable of operating between 300 K and 60 mK. The temperature stabilization, realized through a computer-controlled feedback loop, was better than 0.01 K. The sample temperature was measured with a RuOx resistor thermometer, in contact with the sample holder. The samples were characterized via the use of a Stanford Research Systems SIM921 AC resistance bridge which applied a bias of 3 μA, avoiding any potential self-heating effect. The sample temperature was varied over several hours, allowing time for the complete thermalization of the sample with the sample holder.

The magnetic field transport measurements were carried on a Janis CCS-300S cold finger cryocooler (Janis Research Company, Inc, USA) in a temperature range from 325 K to 10 K. This system is a closed cycle refrigerator based on the Gifford-McMahon thermodynamic cycle coupled with a single stage water cooled He gas compressor (Sumitomo HC-4E1 Helium Compressor, Sumitomo (SHI) Cryogenics of America, Inc., USA). Magnetic characterization is performed by coupling the cryostat and compressor with an electromagnet (Lake Shore Cryotronics Model EM4-HVA, Lakeshore Inc., USA) with induced magnetic fields up to 8400 G. $R(B)$ curves were acquired by sending currents ranging from 0 to ±50 A to the magnet (where ± stands for the current orientation) with a 5 A step; the induced magnetic field intensity is measured with an Hall Probe (425 Gaussmeter, Lake Shore Cryotronics, Inc). The measured maximum hysteresis in the magnetic field is less than 0.01 T throughout the entire electromagnet bias current range, as can be seen in Fig. 7.

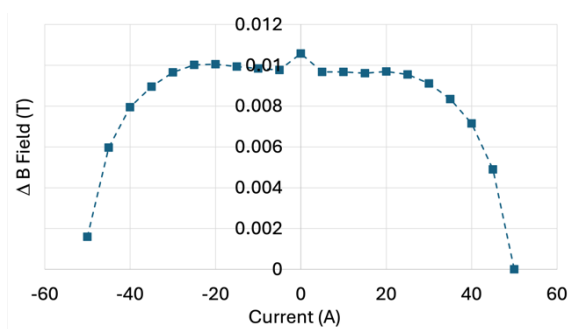

**Figure 7.** Electromagnet hysteresis expressed in variation of magnetic field, measured sweeping the bias of the electromagnet first in one direction and then in the opposite.

## Data Availability

The data that support the findings of this study are available from the corresponding authors upon reasonable request.

## Acknowledgements

The authors would like to thank S. Abate of CNR-SPIN Salerno (Italy) for his technical support and F. Romeo of University of Salerno for fruitful discussions.

This research was funded by Argonne National Laboratory LDRD proposals 2018-002 and 2021-0059; is supported by the Accelerator and Detector R&D program in Basic Energy Sciences' Scientific User Facilities (SUF) Division at the Department of Energy; used resources of the Advanced Photon Source and Center for Nanoscale Materials, U.S. Department of Energy (DOE) Office of Science User Facilities operated for the U.S. DOE, Office of Basic Energy Sciences under Contract No. DE-AC02-06CH11357".

University of Salerno has partially supported this work through grants 300391FRB20BARON, 300391FRB21CAVAL, 300391FRB22PAGAN, 300391FRB23BARON and through the "Visiting Professor 2024" program.

The Project PRIN PNRR 2022 entitled "Development of two-dimensional environmental gas nano-sensors with enhanced selectivity through fluctuation spectroscopy (2DEGAS)" (CUP: D53D23019410001), funded by European Union – Next Generation UE and by Ministry of University and Research (MUR), is also acknowledged for the financial support.

INFN is also gratefully acknowledged through experiments QUB-IT, and DARTWARS.

## Author contributions

O.Q., N.C., A.G., C.B., and S.P. performed the electrical measurements. L.G. prepared the samples and performed the morphological characterization. All authors analyzed and interpreted the data, and equally contributed to the paper writing. The submitted version of the manuscript was agreed by all.

## Competing interests

The authors declare no competing interests.

## Additional information

**Supplementary information** The online version does not contain supplementary material.

**Correspondence** and requests for materials should be addressed to Orlando Quaranta.

**Reprints and permission information** is available at http://www.nature.com/reprints

**Publisher's note** Springer Nature remains neutral with regard to jurisdictional claims in published maps and institutional affiliations.